\documentclass{article}
\usepackage{spconf,amsmath,graphicx}
\usepackage{epstopdf}
\usepackage{cite}
\usepackage{bm}
\usepackage{booktabs}
\usepackage{amsmath}
\usepackage{color}
\usepackage{subcaption}
\usepackage{amssymb}
\usepackage[bottom]{footmisc}
\usepackage{soul}

\def\bfomega{{\bm\omega}}
\def\bfI{{\mathbf{I}}}
\def\bfpi{{\bm\pi}}
\def\bfmu{{\bm\mu}}
\def\bfsigma{{\bm\sigma}}
\def\bfo{\mathbf{o}}
\def\bfO{\mathbf{O}}
\def\bfy{\mathbf{y}}
\def\bfY{\mathbf{Y}}

\title{mixture factorized auto-encoder for unsupervised hierarchical deep factorization of speech signal}
\name{Zhiyuan Peng$^{\dagger}$, Siyuan Feng$^{\dagger}$, Tan Lee$^{\dagger}$
}
\address{$^{\dagger}$Department of Electronic Engineering, The Chinese University of Hong Kong, Hong Kong SAR, China\\
\begin{footnotesize}
\texttt{jerrypeng1937@gmail.com, siyuanfeng@link.cuhk.edu.hk, tanlee@ee.cuhk.edu.hk}
\end{footnotesize}
}

\begin{document}
\ninept
\maketitle
\begin{abstract}

Speech signal is constituted and contributed by various informative factors, such as linguistic content and speaker characteristic. There have been notable recent studies attempting to factorize speech signal into these individual factors without requiring any annotation. These studies typically assume continuous representation for linguistic content, which is not in accordance with general linguistic knowledge and may make the extraction of speaker information less successful. This paper proposes the mixture factorized auto-encoder (mFAE) for unsupervised deep factorization. The encoder part of mFAE comprises a frame tokenizer and an utterance embedder. The frame tokenizer models linguistic content of input speech with a discrete categorical distribution. It performs frame clustering by assigning each frame a soft mixture label. The utterance embedder generates an utterance-level vector representation. A frame decoder serves to reconstruct speech features from the encoders' outputs. The mFAE is evaluated on speaker verification (SV) task and unsupervised subword modeling (USM) task. The SV experiments on VoxCeleb 1 show that the utterance embedder is capable of extracting speaker-discriminative embeddings with performance comparable to a x-vector baseline. The USM experiments on ZeroSpeech 2017 dataset verify that the frame tokenizer is able to capture linguistic content and the utterance embedder can acquire speaker-related information.

\end{abstract}
\section{Introduction}\label{sec:introduction}
Speech signal contains a rich array of information, including linguistic information, e.g., phonetic content, and paralinguistic information, e.g., speaker, emotion, channel distortion, etc. Research on extracting and identifying these information from speech has been abundant. In this paper, the above different aspects of information are regarded as the underlying factors that jointly contribute to the realization of speech.
These informative factors are closely coupled such that they can not be separated using a shallow model with explicit formulations for factorization.
Recently, deep encoders have shown effectiveness in factor extraction through a feature learning process. Examples of these features include linguistic features for unsupervised acoustic unit discovery\cite{chorowski2019unsupervised}, affect-salient features for speech emotion recognition (SER)\cite{mao2014learning}, noise-robust speaker embeddings for speaker recognition (SRE)\cite{snyder2017deep} and phonetically-aware  bottleneck features for language recognition (LRE)\cite{ferrer2015study}.
This motivates investigations on deep factorization of speech signal \cite{li2018deep, Liu2018, NIPS2017_HsuFHVAE, LiM18a, Stafylakis2019}.

One of the representative ideas in realizing deep factorization is to infer multiple speech factors in a sequential manner, i.e.,  factors previously inferred are used as conditional variables in subsequent inference of other factors\cite{li2018deep, Stafylakis2019}.
In \cite{li2018deep}, speaker factor extractor is trained based on the output of a pre-trained ASR system. The multiple-step training strategy is not efficient and may lead to sub-optimal factorization results. A follow-up study adopted multi-task learning to jointly train factor extractors\cite{Liu2018}. This approach requires multiple task-related speech datasets with annotations. 
In speech research, the amount of un-annotated data is much greater than that of annotated ones. To leverage the massive un-annotated data, attempts to unsupervised deep factorization were made \cite{NIPS2017_HsuFHVAE,LiM18a}. There are two major assumptions made in these attempts. Under the first assumption, linguistic information in a speech utterance could be represented by a dynamic factor, while paralinguistic information is represented by a stationary factor. This led a hierarchical design of factor extractors, i.e., having one extractor operating at frame level (frame encoder) to generate the linguistic factor, and the other at utterance level (utterance embedder) to capture the paralinguistic factor.
The second assumption is that the linguistic factor can be described by a continuous representation. 
We argue that this is not in accordance with general linguistic knowledge. Linguistic representation is discrete by nature. Phonetic units are categorically defined. Speech transcriptions are sequences of discrete words or phonetic units. 
In a preliminary experiment, we observed that assuming continuous factor representation in the frame encoder would cause paralinguistic information to be partly absorbed by the frame encoder, making the utterance embedder fail to learn informative representation. It is known as posterior collapse \cite{DBLP:conf/iclr/LucasTGN19},
meaning that the decoder learns to ignore a subset of encoded variables. 
A straightforward solution is 
to add a discriminative loss term \cite{NIPS2017_HsuFHVAE} to obtain a more informative utterance embedder. 
However, this solution requires high computational cost and an additional hyper-parameter for tuning.

\begin{figure}[t]
    \centering
    \includegraphics[width=0.66\linewidth]{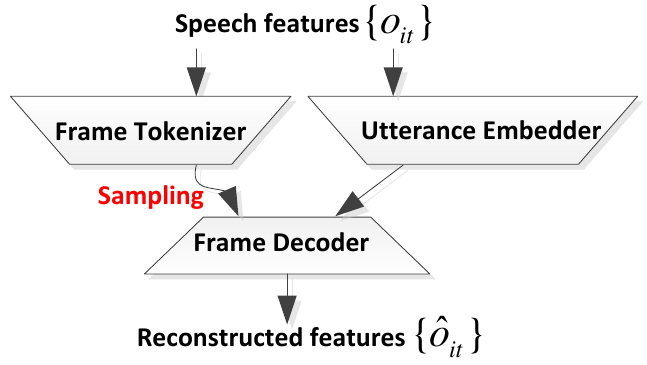}
    \caption{General framework of mixture factorized auto-encoder}
    \label{fig:ufae_framework}
    \vspace{-6mm}
\end{figure}

In the present study, a novel design of mixture factorized auto-encoder (mFAE) is proposed for unsupervised deep factorization. It follows the hierarchical design of factor extractors and assumes discrete representation of linguistic factor. 
As depicted in Figure \ref{fig:ufae_framework}, the mFAE consists of three components: frame tokenizer, utterance embedder and frame decoder. Given a sequence of speech features, the frame tokenizer assigns each frame a soft mixture label (mixture posterior), which constitutes the linguistic factor. 
The utterance embedder generates a vector representation of the whole sequence, which desirably captures the paralinguistic factor.
The frame decoder reconstructs a speech frame from the utterance embedding and the respective frame representation. The frame representation is an \textit{approximation of the one-hot vector} sampled from the mixture posterior. 
The sampling plays the key role in the proposed mFAE. It can be regarded as performing vector quantization of the mixture posterior and thus lays the discrete nature of linguistic factor.
The three components are jointly trained with the mean-squared error loss.

The mFAE is a simplified variant of mixture factorized variational auto-encoder (mFVAE). The thoughts behind the simplifications are rich as we will explain in the next section.


\section{mixture factorized auto-encoder}




\subsection{Formulation of mFVAE}
\begin{figure}[th]
\begin{subfigure}{.5\linewidth}
  \centering
  \includegraphics[width=1.0\linewidth]{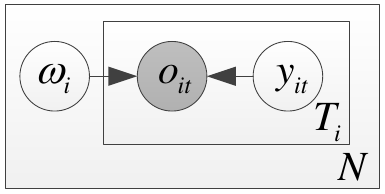}  
  \caption{Generative model\footnotemark[1]}
  \label{fig:generativemodel}
\end{subfigure}
\begin{subfigure}{.5\linewidth}
  \centering
  \includegraphics[width=1.0\linewidth]{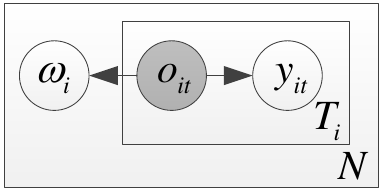}  
  \caption{Inference model}
  \label{fig:inferencemodel}
\end{subfigure}
\caption{Graphical illustration of the proposed mFVAE. 
The shaded circles represent observed speech features $o_{it}$. The white circles represent latent variables.
}
\label{fig:fig}
\end{figure}

\footnotetext[1]{The generative model is almost the same as that of i-vector method \cite{DBLP:journals/taslp/DehakKDDO11}, except that $p(\bfo_{it}|\bfomega_i, \bfy_{it})$ is parameterized by a neural network to characterize the non-linear relationship between $\bfomega_i$ and $\bfy_{it}$.}
\footnotetext[2]{The uniform regularization refers to $D_{KL}(q(\bfy_{it})||p(\bfy_{it}))$. It can be proved that $\mathbb{E}_{p(\bfo_{it})} \left[ D_{KL}(q(\bfy_{it}|\bfo_{it})||p(\bfy_{it})) \right] \geq D_{KL}(q(\bfy_{it})||p(\bfy_{it}))$ and $\min\mathbb{E}_{p(\bfo_{it})} \left[ D_{KL}(q(\bfy_{it}|\bfo_{it})||p(\bfy_{it})) \right] = \min KL(q(\bfy_{it})||p(\bfy_{it}))$.}

Given a set of $N$ i.i.d speech utterances $\mathcal{D}=\{\bfO_{i}\}_{i=1}^{N}$, where the $i$-th utterance $\bfO_{i}=\{\bfo_{it}\}_{t=1}^{T_i}$ consists of $T_i$ frames, the mFVAE assumes that $\bfO_{i}$ is generated by the following process:
\begin{enumerate}
    \item A \textbf{sequence vector} $\bfomega_i$ is generated from a Gaussian distribution $p(\bfomega) = \mathcal{N}(\bfomega | \mathbf{0}, \bfI)$.
    \item A sequence of i.i.d. one-hot \textbf{mixture indicators} \ $\bfY_i=\{\bfy_{it}\}_{t=1}^{T_i}$ is drawn from a Categorical distribution $p(\mathbf{y}) = \text{Cat} (\bfpi)$, where $\bfpi=\{\pi_k = \frac{1}{K}\}_{k=1}^{K}$.
    \item Each frame $\bfo_{it}$ in $\bfO_i$ is generated from $p(\mathbf{o} | \bfomega_i, \bfy_{it}) = \mathcal{N}(\mathbf{o} | \bfmu_\mathbf{o}(\bfomega_i, \bfy_{it}), \bfI)$.
\end{enumerate}
The generative process is illustrated in Figure \ref{fig:generativemodel}. The joint probability of $\{\bfO_i, \bfY_i, \bfomega_i\}$ is factorized as,
\begin{equation}
    p(\bfO_i, \bfY_i, \bfomega_i) =  p(\bfomega_i)\prod_{t=1}^{T_i} p(\bfy_{it})p(\bfo_{it}|\bfomega_i,\bfy_{it}).
\end{equation}
Similar to the VAE framework \cite{DBLP:journals/corr/KingmaW13}, the mFVAE requires an inference model $q(\bfomega_i, \bfY_i | \bfO_i)$ to approximate the intractable posterior $p(\bfomega_i, \bfY_i | \bfO_i)$. 
We consider the following inference model as illustrated in Figure \ref{fig:inferencemodel},
\begin{equation}
    q(\bfomega_i, \bfY_i | \bfO_i) =  q(\bfomega_i | \bfO_i)\prod_{t=1}^{T_i} q(\bfy_{it}|\bfo_{it}),
\end{equation}
where 
$q(\bfomega_i|\bfO_i) = \mathcal{N}(\bfomega_i | \bfmu_{\bfomega}(\bfO_i), \bfsigma_{\bfomega}^2 (\bfO_i))$ and $q(\bfy_{it}|\bfo_{it})=\text{Cat}(\bfpi_\mathbf{y}(\bfo_{it}))$.
The functions $\{ \bfmu_{\mathbf{o}}(\cdot, \cdot), \bfmu_{\bfomega}(\cdot), \bfsigma_{\bfomega}^2 (\cdot), \bfpi_\mathbf{y}(\cdot)\}$ are all parameterized by neural networks. 
Similar to the Joint-VAE framework\cite{dupont2018learning}, the loss function for training mFVAE is the negative variational lower-bound,
\begin{equation}\label{equ:mfvae_loss}
\begin{split}
     -\mathcal{L_\text{mFVAE}} (\mathcal{D}) = &\sum_{i=1}^{N} \mathbb{E}_{q(\bfomega_i, \bfY_i | \bfO_i)}\left[ -\log p(\bfO_i|\bfomega_i, \bfY_i) \right] \\
    &+ \beta_\bfomega \sum_{i=1}^{N} D_{KL}\left[ q(\bfomega_i| \bfO_i) || p(\bfomega_i)\right] \\
    &+ \beta_\bfy \sum_{i=1}^{N}\sum_{t=1}^{T_i} D_{KL}\left[ q(\bfy_{it}|\bfo_{it}) || p(\bfy_{it}) \right].
\end{split}
\end{equation}

\begin{figure*}[t!]
    \centering
    \includegraphics[width=0.65\linewidth]{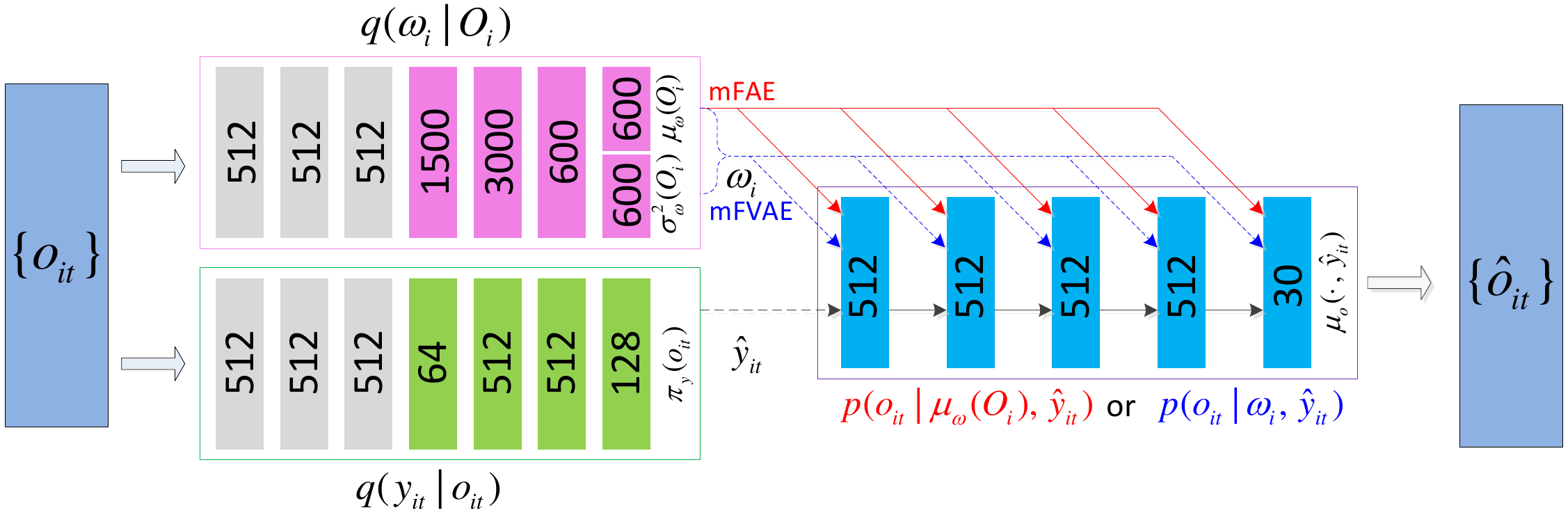}
    \caption{Network structure of mFVAE/mFAE. The numbers are the layer output dimensions. The dashed lines represent sampling $\hat{\bfy}_{it}$ and $\bfomega_i$ from distribution parameterized by neural networks. The frame decoder is  $p(\bfo_{it}|\bfmu_\bfomega(\bfO_i), \hat{\bfy}_{it})$ for mFAE or $p(\bfo_{it}|\bfomega_i, \hat{\bfy}_{it})$ for mFVAE.}
    \label{fig:networkconfig}
    \vspace{-5mm}
\end{figure*}

\subsection{From mFVAE to mFAE}

\subsubsection{Discard regularization losses}

The loss function $ -\mathcal{L_\text{mFVAE}} (\mathcal{D})$ has two components: the reconstruction loss $\mathbb{E}_{q(\bfomega_i, \bfY_i | \bfO_i)}\left[ -\log p(\bfO_i|\bfomega_i, \bfY_i) \right]$ and the factor regularization losses $D_{KL}\left[ q(\bfomega_i| \bfO_i) || p(\bfomega_i)\right], D_{KL}\left[ q(\bfy_{it}|\bfo_{it}) || p(\bfy_{it}) \right]$. 
The reconstruction loss measures the expectation of $L2$ distance between the input speech feature $\bfo_{it}$ and the reconstructed one $\hat{\bfo}_{it}$ over the latent variables $ \bfomega_i, \bfy_{it}$. With this loss, the encoders $q(\bfy_{it} | \bfo_{it})$ and $q(\bfomega_i | \bfO_i)$ are forced to encode linguistic (frame level) and paralinguistic factors (utterance level) respectively.
The factor regularization losses measure the KL distances between the informative posteriors $q(\bfomega_i|\bfO_i), q(\bfy_{it}|\bfo_{it})$ and the non-informative priors $p(\bfomega_i), p(\bfy_{it})$. With these losses, the latent variables are forced to conform the following priors:
\begin{enumerate}
    \item[Prior 1] $\bfomega_i$ follows \textit{standard} Gaussian distribution;
    \item[Prior 2] $\bfy_{it}$ follows \textit{uniform} Categorical distribution.
\end{enumerate}

As justified in the next two paragraphs, these two priors are incompatible with the intended function of mFVAE: to factorize speech into linguistic factor and paralinguistic factor.
We propose to remove the regularization losses so as to get rid of the two incompatible priors. 

For Prior 1: (1) Standard Gaussian distribution assumes the independence between the dimensions of sequence vector $\bfomega_i$. 
This assumption is unnecessary for $\bfomega_i$ to capture paralinguistic information;
(2) Requiring $\bfomega_i$ to conform Gaussian distribution may be beneficial to weak back-end classifiers. But a recent study\cite{Zhang2019} showed that the benefit vanishes when using strong discriminant back-ends like linear discriminant analysis (LDA) incorporated with Probabilistic LDA (PLDA);
(3) The first regularization loss makes $\bfomega_i$ less informative, which is against our goal;
(4) A concern about removing the regularization loss is that $\bfomega_i$ may contain dynamic linguistic information, making the frame decoder ignore $\bfy_{it}$. This issue could be solved by using a feed-forward neural network as the frame decoder, because a feed-forward neural network is unable to reconstruct the whole utterance from only a sequence vector $\bfomega_i$.


For Prior 2: (1) The second regularization loss serves as a soft constraint to prevent $q(\bfy_{it}|\bfo_{it})$ being deviating too much from $p(\bfy_{it})$ and in this way it constraints the information $\bfy_{it}$ conveyed to the decoder $p(\bfo_{it}|\bfomega_i, \bfy_{it})$. In our model, $\bfy_{it}$ is a one-hot vector so that the amount of information being conveyed is constrained to be less than $\log_2(K)$ bit. 
(2) A potential problem of discarding the regularization loss is that the uniform regularization of the mixture weight vector $q(\bfy_{it})$ is also removed\footnotemark[2]. 
This uniform regularization turns out to maximize the average amount of information (entropy) being conveyed by $q(\bfy_{it}|\bfo_{it})$. Keeping this uniform regularization may encourage a more informative $q(\bfy_{it}|\bfo_{it})$. However, no noticeable benefit to factorization is found in our preliminary experiments.



\subsubsection{Replace $\mathcal{N}(\bfomega_i | \bfmu_\bfomega(\bfO_i), \bfsigma_{\bfomega}^2(\bfO_i))$ with $\delta_{\bfmu_\bfomega(\bfO_i)} (\bfomega_i)$}\label{sec:simpOfomega}

$\bfomega_i$ can be represented as $\bfmu_\bfomega(\bfO_i)$ being corrupted by a Gaussian noise $\mathcal{N}(\mathbf{0}, \bfsigma_{\bfomega}^2(\bfO_i))$. The noise variance $\bfsigma_\bfomega^2(\bfO_i)$ parameterized by neural network tends to $\mathbf{0}$ when only the reconstruction loss is being minimized. This trend is resisted by heavy penalization of the factor regularization term $D_{KL}\left[ q(\bfomega_i| \bfO_i) || p(\bfomega_i)\right]$ around $\bfsigma_\bfomega^2(\bfO_i) = \mathbf{0}$. As a consequence of discarding the regularization term, the posterior $\mathcal{N}(\bfmu_\bfomega(\bfO_i), \bfsigma_{\bfomega}^2(\bfO_i))$ becomes spiky. We use an impulse $\delta_{\bfmu_\bfomega(\bfO_i)} (\bfomega_i)$ to approximate it. Thus the reconstruction loss can be simplified as $\mathcal{L}_{\text{sim\_rec}}$ = $\mathbb{E}_{q(\bfY_{i}|\bfO_{i})}\left[ -\log p(\bfO_i| \bfmu_\bfomega(\bfO_i), \bfY_i) \right]$.





\subsubsection{Apply reparameterization for $\text{Cat}(\bfpi_{\mathbf{y}}{(\bfo_{it})})$}
The simplified reconstruction loss $\mathcal{L}_{\text{sim\_rec}}$ is computational tractable for mini-batch stochastic gradient descent.
Its computation complexity is $O(BLK^2)$ where $B$ denotes the number of speech utterances in each batch and $L$ is the average number of frames per utterance. We  attempt to reduce the computation complexity to $O(BLK)$ in order to speed up the training on large-scale datasets.





Sampling $\bfy_{it}$ from $q(\bfy_{it}|\bfo_{it})$ is easy. A typical solution is to apply the Gumbel-Max trick\cite{NIPS2014_5449},
\begin{equation}
    \bfy_{it} = \text{one\_hot} \left(\operatorname*{argmax}_k \left[ \log(\bfpi_\mathbf{y}(\bfo_{it})) + \bm{g} \right]\right),
\end{equation}
where $\bm{g} = [g_1,\cdots,g_K]$.  $g_1,\cdots,g_K$ are i.i.d. samples drawn from Gumbel$(0,1)$. 
The problem here is how to make the one-hot vector $\bfy_{it}$ differentiable for gradient back-propagation through $\bfpi_\mathbf{y}(\bfo_{it})$. We adopt the Gumbel-Softmax distribution to draw differentiable sample $\hat{\bfy}_{it}$ to approximate $\bfy_{it}$ as proposed in \cite{DBLP:conf/iclr/JangGP17},
\begin{equation}\label{equ:gumbelSoftmax}
\hat{\bfy}_{it} = \text{softmax} ( (\log(\bfpi_\mathbf{y}(\bfo_{it})) + \bm{g})/\tau),
\end{equation}
where the softmax temperature $\tau$ is a positive hyper-parameter. The choice of $\tau$ is a trade-off between the variance of reparameterization gradients and the approximation of $\bfy_{it}$. A large $\tau$ provides smooth $\hat{\bfy}_{it}$, poor approximation of $\bfy_{it}$, but small variance of the gradients. Whilst a small $\tau$ produces $\hat{\bfy}_{it}$ close to one-hot but large variance of the gradients. As $\tau \to 0$, $q(\hat{\bfy}_{it}|\bfo_{it}) \to \text{Cat}(\bfpi_\mathbf{y}(\bfo_{it})) = q(\bfy_{it}|\bfo_{it})$.

\bigskip
In summary, the loss function of mFAE is simplified as,
\begin{equation}\label{equ:mfae_loss}
    \begin{split}
        -\mathcal{L}_{\text{mFAE}}(\mathcal{D}) &= \sum_{i=1}^{N} \mathbb{E}_{q(\bfomega_i, \bfY_i | \bfO_i)} \left[ -\log p(\bfO_i | \bfomega_i, \bfY_i)  \right] \\
        &\approx \sum_{i=1}^N \mathbb{E}_{q(\bfY_i|\bfO_i)\delta_{\bfmu_\bfomega{(\bfO_i)}}} \left[ -\log p(\bfO_i | \bfomega_i, \bfY_i)  \right] \\
        &= \sum_{i=1}^N\sum_{t=1}^{T_i} \mathbb{E}_{q(\bfy_{it}|\bfo_{it})} \left[ -\log p(\bfo_{it}|\bfmu_\bfomega(\bfO_i), \bfy_{it})\right]\\
        &\approx \sum_{i=1}^N\sum_{t=1}^{T_i} -\log p(\bfo_{it}|\bfmu_\bfomega(\bfO_i),\hat{\bfy}_{it})\\
        &= \frac{1}{2}\sum_{i=1}^N\sum_{t=1}^{T_i} ||\bfo_{it}-\bfmu_{\mathbf{o}}(\bfmu_\bfomega(\bfO_i), \hat{\bfy}_{it}) ||^2_2 + \text{const},
    \end{split}
\end{equation}
where the trainable parameters are $\{\bfmu_\mathbf{o}(\cdot, \cdot), \bfmu_\bfomega(\cdot), \bfpi_\mathbf{y}(\cdot)\}$. $\hat{\bfy}_{it}$ is generated by Equation (\ref{equ:gumbelSoftmax}). 
$\tau$ is set to $0.1$ in our experiments.

\subsection{Network structure of mFVAE/mFAE}

The network structure of mFVAE/mFAE is implemented as shown in Figure \ref{fig:networkconfig}.
It consists of three components: utterance embedder $q(\bfomega_i|\bfO_i)$, frame tokenizer $q(\bfy_{it}|\bfo_{it})$ and frame decoder $p(\bfo_{it}|\cdot, \hat{\bfy}_{it})$. The utterance embedder consists of four TDNN layers with contexts of $[-2,-2], \{-2, 2\}, \{-3, -3\}, \{0\}$, a mean+std global pooling layer across time\cite{snyder2018x}, two feed-forward layers, a linear output layer parameterizing $\bfmu_\bfomega(\bfO_i)$ and a softplus output layer parameterizing $\bfsigma_\bfomega^2(\bfO_i)$.
The frame tokenizer has four TDNN layers with the same context as those in the utterance embedder, followed by two feed-forward layers and a softmax layer that parameterizes $\bfpi_\mathbf{y}(\bfo_{it})$. 
The frame decoder has a TDNN input layer with a context of $[-1, 1]$, followed by a four-layer feed-forward neural network. 
The sequence vector ($\bfomega_i$ for mFVAE and $\bfmu_\bfomega(\bfO_i)$ for mFAE) is appended to the input of each layer of $p(\bfo_{it}|\cdot, \hat{\bfy}_{it})$. Its output layer parameterizes $\bfmu_\mathbf{o}(\cdot, \hat{\bfy}_{it})$.
All layers except output layers in the three components are followed by ReLU and batch normalization.

\section{Experiments}




The proposed approach of deep factorization is evaluated on two tasks, namely, speaker verification and unsupervised subword modeling. The speaker verification task is adopted for evaluating speaker identity, a major part of paralinguistic factor, from $q(\bfomega_i|\bfO_{i})$. The task of unsupervised subword modeling is chosen to explore the phonetic information, a type of linguistic factor, from $q(\bfy_{it}|\bfo_{it})$. 

\subsection{Speaker verification on VoxCeleb 1}\label{sec:exp_voxceleb}

VoxCeleb 1 is a large-scale public speech corpus\cite{nagrani2017voxceleb}. The training set consists of $148,642$ utterances from 1211 speakers, with a total speech duration of about $300$ hours. The test set has $4,874$ utterances from $40$ speakers. 
The performance of speaker verification is evaluated in terms of equal error rate (EER) and the minimum of normalized detection cost function (mDCF). 
EER refers to the rate at which both false acceptance rate and false rejection rate are equal. mDCF is a weighted sum of false acceptance rate and false rejection rate that usually penalizes more on false acceptance rate.


30-dimensional MFCCs (without cepstrum truncation) with global mean variance normalization (GMVN) are used as input features for mFVAE and mFAE. The loss function is given by Equation (\ref{equ:mfvae_loss}) for mFVAE or by Equation (\ref{equ:mfae_loss}) for mFAE. We use the ADAM optimizer for training, with a learning rate exponentially decreased from 1e-3 to 1e-4 in 50 epochs. Each batch consists of 64 randomly-trimmed segments of 3-second long.

After training, utterance embeddings $\bfmu_\bfomega(\bfO_i)$ are extracted, preprocessed by LDA to reduce the dimension from 600 to 150, followed by length-normalization and a two-covariance PLDA\cite{sizov2014unifying} classifier for similarity scoring between embeddings. The evaluation metrics, EER and mDCF, are computed based on these scores. 



The i-vector\cite{DBLP:journals/taslp/DehakKDDO11} and x-vector\cite{snyder2018x} baselines are established for performance comparison. The i-vector front-end includes a 2048-mixture GMM-UBM model and a 600-dim i-vector extractor. The x-vector front-end follows the standard Kaldi x-vector architecture. Their back-ends are the same as those of mFAE and mFVAE.


\begin{table}[h]
\renewcommand\arraystretch{0.5}
\caption{EER\% and mDCF with P(tar)=0.01 on VoxCeleb 1. For simplicity, there is no data augmentation in all experiments.}
\label{tab:voxceleb1}
\centering
\begin{tabular}{llll}
\toprule
 & & EER\% & mDCF \\
 \toprule[1pt]\midrule[0.3pt]
mFVAE, $\beta_\bfy=0$     & $\beta_\bfomega=3$    & $11.48$ & $0.761$\\
         & $\beta_\bfomega=1$    & $9.42$  & $0.698$ \\
         & $\beta_\bfomega=0.01$ & $7.20$ & $0.595$  \\
\midrule
mFVAE, $\beta_\bfomega=0$ & $\beta_\bfy=1$ & $19.40$ & $0.892$ \\
         & $\beta_\bfy=0.1$ & $9.17$ & $0.697$\\
         & $\beta_\bfy=0.01$ & $7.25$ & $0.579 $ \\
\midrule
mFAE      &         & $7.39$ & $0.589$ \\
x-vector &         & $7.49$ & $0.650$ \\
i-vector &         & $\mathbf{5.51}$ & $\mathbf{0.462}$ \\
 \toprule[1pt]\bottomrule[0.3pt]
\end{tabular}
\end{table}

The results in Table \ref{tab:voxceleb1} show that: (1) For mFVAE with $\beta_\bfy = 0$, the decrease of $\beta_\bfomega$ leads to consistent performance improvement in terms of both EER and mDCF. This indicates that with less penalty on $\bfomega_i$, the utterance embedding contains richer speaker information. (2) The performance improvement is more obvious for mFVAE with $\beta_\bfomega = 0$ as $\beta_\bfy$ decreases. A high penalty on $\beta_\bfy$ stops the frame decoder to extract dynamic linguistic information from $\bfy_{it}$. This leads to a trivial solution that the reconstructed feature $\bfo_{it}$ equals to the mean vector of the $i$-th utterance and $\bfomega_i$ fails to capture speaker information. (3) The proposed mFAE is capable of extracting speaker-discriminant embeddings with performance comparable to the x-vector method that requires speaker labels for supervision. (4) The i-vector method performs the best on this dataset. It is known that neural network approaches are data-hungry. With data augmentation\cite{DBLP:conf/odyssey/SnyderGMSPK18} and larger datasets like VoxCeleb 2\cite{chung2018voxceleb2}, neural network approaches achieve better performance than the i-vector method. Nevertheless, for applications with limited training data, i-vector warrants in-depth investigation. This classical method may serve as a guidance for the design of neural network models.


\subsection{Unsupervised subword modeling on ZeroSpeech 2017}

The goal of ZeroSpeech 2017 Track 1 is to construct a frame-wise feature representation of speech sounds,
which is both phoneme-discriminative and speaker-invariant, under the assumption that only untranscribed data are available. The dataset consists of three languages, namely, English, French and Mandarin. The amount of training data for the three languages are 45, 24 and 2.5 hours respectively. 
The subset of test data with 10-second segment length is adopted in our study. 
The ABX error rate is used as the evaluation metric for both within-/across-speaker conditions\cite{dunbar2017zero}. 

The input features to mFAE are 13-dimension MFCCs with the first and the second-order time derivatives followed by cepstral mean normalization (CMN) with 3-second sliding window and GMVN are fed into mFAE. The feature vector is denoted as \textit{MFCC+GMVN}. The training process is the same as mentioned in Section \ref{sec:exp_voxceleb}.


After the mFAE training, the reconstructed feature $\hat{\bfo}_{it}$ is obtained from the frame decoder conditioned on two settings: (1) per-utt: $\bfpi_\bfy(\bfo_{it})$ and $\bfmu_\bfomega(\bfO_i)$; (2) unified: $\bfpi_\bfy(\bfo_{it})$ and 
$\overline\bfmu_\bfomega = \frac{1}{|\mathcal{D}|}\sum_{\bfO_j\in \mathcal{D}} \bfmu_\bfomega(\bfO_j)$. Here $\mathcal{D}$ denotes the training set.

The recommended baseline set-up is to evaluate the ABX error rate on MFCC features post-processed by CMN\cite{chorowski2019unsupervised}. 
Here we provide an additional baseline using the feature MFCC+GMVN to allow a fair comparison with mFAE.

\begin{table}[h]
\renewcommand\arraystretch{0.89}
\centering
\caption{ABX\% error rates on ZeroSpeech 2017. mFAE is trained separately on three languages.}
\resizebox{1 \linewidth}{!}{%
\begin{tabular}{l|cc|cc|cc}      
 \toprule[1pt]\midrule[0.3pt]
& \multicolumn{2}{c|}{English 10s} & \multicolumn{2}{c|}{French 10s} & \multicolumn{2}{c}{Mandarin 10s}\\
 & within-spk & across-spk & within-spk & across-spk & within-spk & across-spk \\
\midrule
MFCC & $10.52$ & $20.83$ & $11.19$ & $23.08$ & $9.66$ & $15.39$ \\
~+ GMVN           & $\bm{8.16}$ & $18.95$ & $\bm{10.23}$ & $21.74$ & $\bm{8.76}$ & $\bm{14.92}$ \\
mFAE, per-utt & $10.00$ & $16.80$ & $12.89$ & $19.26$ & $14.06$ & $16.86$ \\
mFAE, unified & $9.88$ & $\bm{15.21}$ & $12.66$ & $\bm{18.03}$ & $13.94$ & $15.62$ \\
\midrule[0.3pt]\bottomrule[1pt]
\end{tabular}
}
\label{tab:zr17result}
\end{table}

The ABX error rates on both input MFCC and reconstructed MFCC across three languages are shown in Table \ref{tab:zr17result}. It is found that: 
(1) In within-speaker condition, features reconstructed by mFAE show slight performance degradation as compared to MFCC + GMVN. 
A potential problem is the insufficient training of mFAE as the number of epochs is fixed to 50 for all of the three language. The degradation is more noticeable on Mandarin, which only has $2.5$ hours of training data. 
(2) As $\overline{\bfmu}_\bfomega$ is computed from the training set, it doesn't contain the dynamic linguistic information in the test utterances. Thus, the performance of (\textit{mFAE, unified}) in within-/across-speaker conditions proves that $q(\bfy_{it}|\bfo_{it})$ captures linguistic information.
(3) The performance improves consistently in the across-speaker condition, comparing (\textit{mFAE, per-utt}) with (\textit{mFAE, unified}). This shows that $q(\bfomega_i|\bfO_{i})$ is capable of extracting paralinguistic information.


\section{Conclusions}
In this paper, we propose mixture factorized auto-encoder (mFAE), a scalable unsupervised hierarchical deep factorization approach to decomposing speech into paralinguistic factor and linguistic factor. The two factors are represented by a sequence vector and a frame-wise mixture indicator respectively. Experiments on speaker verification and unsupervised subword modeling show that the sequence vector contains rich speaker information and mixture indicator keeps linguistic content.
mFAE can be applied to many down-stream tasks, like text-to-speech and query-by-example spoken term detection.

\bibliographystyle{IEEEtran}
\begin{footnotesize}
\bibliography{refs,strings}
\end{footnotesize}

\end{document}